\documentstyle[12pt]{article}

\begin{document}
\setcounter{page}{1} \pagestyle{plain} \vspace{1cm}
\begin{center}
\Large{\bf Rip Singularity Scenario and Bouncing Universe in a Chaplygin Gas Dark Energy Model}\\
\small \vspace{1cm} {\bf S. Davood
Sadatian}\footnote{sd-sadatian@um.ac.ir}\\
\vspace{0.5cm} {\it Department of Physics,
Faculty of Basic Sciences,\\
University of Neyshabur,\\
P. O. Box 91136-899, Neyshabur, Iran
 }\\
\end{center}
\vspace{1.5cm}
\begin{abstract}
We choose a modified Chaplygin Gas Dark energy model for considering
some its cosmological behaviors. In this regards, we study different
Rip singularity scenarios and bouncing model of the universe in
context of this model. We show that by using suitable parameters can
explain some cosmological aspects of the model.

\end{abstract}
\newpage

\section{Introduction}
There is growing evidence that the Universe at present era is
dominated with a component by negative pressure, known as dark
energy, leading to accelerated expansion for Universe. While the
best candidate for this component is vacuum energy, a conceivable
alternative is dynamical relation for vacuum energy [1,2]. It has
been encouraged that the change of manner of the dark energy density
could be managed with the change in the equation of state (EoS) of
the background fluid as a replacement for the form of the potential,
by means of that avoiding fine-tuning problems. This is obtained via
an unusual background fluid that so-called the Chaplygin gas.
Besides, the Chaplygin gas is the only fluid known to accept a supersymmetric generalization [3].\\
Before results of Supernova data, it seems that the Universe maybe
filled with energy density which is scattered over large scales [4].
The different cosmological observations such as Cosmic Background
Explorer data (COBE) [5,6,7], SNeIa data [8], large scale red-shift
surveys [9], the evaluations of the cosmic microwave background
(CMB) [10] and WMAP data [11] predict the Universe which is
expanding with acceleration in present era. However, for a good
consistency with data, A generalized Chaplygin Gas model has been proposed that known as (MCG) [12].\\
In other view point, according to variation of energy density and
scale factor with time[13], Rip singularity scenario and other cases
of singularity maybe appear in Modified Chaplygin Gas model. These
solutions occur when $\omega< -1$ increases rapidly. Moreover, it is
possible some types of singularity, depend on energy density and
scale factor how increases with time[13]. Also an interesting
solution of the singularity problem in the standard Big Bang
cosmology is Bouncing Universe. A bouncing universe has an initial
narrow state by a minimal radius and then develops to an expanding
phase [14]. This means, for the universe arriving to the Big Bang
era after the bouncing, the equation of state parameter should
crossing from $\omega < -1$ to $ \omega>- 1$. In a bouncing
cosmology, what we comment to be "The Big Bang scenario" really is
the Universe emerging from a bounce. The universe at this position
has its smallest extent (smallest scale factor $a$) and largest
energy density. However, we know that the cosmological model
avoiding the big bang singularity within the frame of Einstein
gravity has to satisfy the effective equation of state less than -1
around the bounce and then enters into regular expansion with
equation of state larger than -1. This point of view was the
associated scenario is usually dubbed as the Quintom scenario as
comprehensively reviewed in [15]. Moreover, the idea of connecting
this scenario with bouncing cosmologies was originally carried out
in [16], and its huge ideas can even be tracked backward to the
cosmic duality analysis in [17]. The latest study on this respect
can be found in the construction of Lee-Wick cosmology in
[18] and known as ``new matter bounce" cosmology in [19].\\
Therefore, according above discussions, we study a modified
Chaplygin Gas in section 1. In section 2, we briefly discuss about
general Dark Energy Model and in two last sections, we consider Rip
Singularity scenario and Bouncing Universe solution in context of
our model. In section 5, we summarize our results and conclusion.
\section{The Modified Chaplygin Gas Model}
In following, according [20], we explain a modified Chaplygin gas in
Friedmann-Robertson-Walker cosmology. This model based on a equation
of state as
\begin{eqnarray}
p & = & A \rho - \frac{B}{\rho^{\alpha}}
\end{eqnarray}
where  $A, B$ and $\alpha$ are constant parameters and usually constraint with observation data. \\
If $B = 0$, we obtain usual equation of state of perfect fluid, when
$A=0$, it products the generalized Chaplygin gas. \\
The metric of $D$-dimensional FRW space-time given as
\begin{eqnarray}
d s^2 & = & - d t^2 + a{^2}(t)~ d \Omega_k{^2}
\end{eqnarray}
where $a (t)$ is the scale factor and $d \Omega_k{^2}$ is the metric
$( D - 1)$--space by curvature $k = 0,\pm 1$ (following we choose D=4). \\
One can obtain the Friedmann equation as
\begin{eqnarray}
H^2 & = & \left( \frac{\dot{a}}{a} \right)^2~ =~ \frac{2~\rho}{(D-1)
(D-2)} - \frac{k}{a^2}
\end{eqnarray}
where $H$ is the Hubble parameter. \\
Now the conservation law for a fluid with an energy density $\rho$
and a pressure $p$ can write as
\begin{eqnarray}
\dot{\rho} + (D-1) H ( \rho + p ) & = & 0.
\end{eqnarray}
By using equations (3) and (4) we have
\begin{eqnarray}
\frac{\ddot{a}}{a} & = & \dot{H} + H^2~ =~ -~ \frac{(D-1)~p +
(D-3)~\rho}{(D-1)(D-2)} ~~~~.
\end{eqnarray}
If we assume $\Gamma = a^{(D - 1) (A + 1)}$ and rewrite density
$\bar{\rho} = \rho \Gamma$, the equation (4) gives
\begin{eqnarray}
\dot{\bar{\rho}} - \frac{B}{A + 1}~\frac{\Gamma^{\alpha}
\dot{\Gamma}}{\bar{\rho}^{\alpha}} & = & 0 ~~.
\end{eqnarray}
Now with integrated from equation (6), we have
\begin{eqnarray}
\frac{\bar{\rho}^{\alpha+1}}{\alpha + 1} & = & \frac{B}{A+1}
\frac{\Gamma^{\alpha+1}}{\alpha+1}~ +~ \frac{C}{\alpha+1}
\end{eqnarray}
where $C$ is an integration constant . \\
The energy density will obtain
\begin{eqnarray} \rho & = & \left(
\frac{B}{A+1} + \frac{C}{\Gamma^{\alpha+1}}
\right)^{\frac{1}{\alpha+1}} ~~~.
\end{eqnarray}
If $C$ expressed in terms of the cosmological scale $a_0$,~( in form
of $\Gamma_0 = a_0^{(D -1 )(A+1)}$~) and when the fluid has a
vanishing pressure, therefore we obtain this parameter in form
\begin{eqnarray}
C & = & \frac{B}{A+1}~ \frac{\Gamma_0^{\alpha+1}}{A}.
\end{eqnarray}
Now we rewrite the energy density $\rho$ as
\begin{eqnarray}
\rho & = & \left( \frac{B}{A+1} \right)^{\frac{1}{\alpha+1}}~ \left(
1 + \frac{1}{A~\Gamma_r^{\alpha+1}}  \right)^{\frac{1}{\alpha+1}}
\end{eqnarray}
where $\Gamma_r = \Gamma/\Gamma_0$. \\
Here, one can take some limits for parameter $\Gamma$ and obtain
interesting cosmological results[20].\\
\subsection{Modified Chaplygin Gas as a Scalar Field}
Following [21,22], we study the modified Chaplygin gas cosmological
model by introducing a scalar fields with a potential $U(\varphi)$
and the Lagrangian in form of
\begin{eqnarray}
{\cal L}_{\varphi} & = & \frac{\dot{\varphi}^2}{2} - U(\varphi) ~~.
\end{eqnarray}
We notice both the energy density and the pressure of the modified
Chaplygin gas depend on the scalar $\varphi$ in the following
transformation equations
\begin{eqnarray}
\rho_{\varphi} & = & \frac{\dot{\varphi}^2}{2} + U(\varphi) =~ \rho  \nonumber \\
p_{\varphi} & = & \frac{\dot{\varphi}^2}{2} - U(\varphi) =~ A \rho -
\frac{B}{\rho^\alpha} ~~~~.
\end{eqnarray}
The kinetic energy of the scalar field given as
\begin{eqnarray}
\dot{\varphi}^2 & = & ( 1 + \omega_{\varphi} )~ \rho_{\varphi}  \nonumber \\
U(\varphi) & = & \frac{1}{2}~( 1 - \omega_{\varphi} )~\rho_{\varphi}
\end{eqnarray}
where $\omega_{\varphi} = p_{\varphi}/\rho_{\varphi}$ ~. \\
We know $\dot{\varphi} = \varphi' \dot{\Gamma_r}$ where the prime
denotes derivation with respect to $\Gamma_r$ and $\dot{\Gamma_r} =
(D -1) (A + 1) H \Gamma_r$, hence we have
\begin{eqnarray}
\varphi'{^2} & = & \frac{D -2}{2 (D -1) (A+1)^2}~\frac{1 +
\omega_{\varphi}}{\Gamma_r^2} ~~.
\end{eqnarray}
However, if we use equation (3) for a flat universe, $k = 0$, we
have
\begin{eqnarray}
\varphi' & = & \sqrt{\frac{D -2}{2 (D -1) A (A +1)}}~
\frac{1}{\Gamma_r^\frac{\alpha+3}{2} \sqrt{1 + \frac{1}{A \Gamma_r^{\alpha+1}}}} \nonumber \\
U \left( \varphi \right) & = & \frac{1}{2} \left(\frac{B}{A+1}
\right)^{\frac{1}{2}}~ \frac{2 + \frac{1 -A}{A
\Gamma_r^{\alpha+1}}}{\sqrt{1 + \frac{1}{A \Gamma_r^{\alpha+1}}} }
~~.
\end{eqnarray}
With integrated from the first equation, we have
\begin{eqnarray}
A \Gamma_r^{\alpha+1} & = & \frac{1}{ \sinh^2 \left( \xi (\alpha+1)
\Delta \varphi \right)}
\end{eqnarray}
where $\xi = \sqrt{\frac{(D-1)(A+1)}{2 (D-2)}}$ and $\Delta \varphi = \varphi - \varphi_0$. \\
Now by using the above relations in equation (12), we can obtain
energy density , pressure $\rho_{\varphi}, p_{\varphi}$ and equation
of state relation $\omega_{\varphi}$ in terms of the scalar field
$\varphi$ as
\begin{eqnarray}
\rho_{\varphi} & = & \left( \frac{B}{A + 1}
\right)^{\frac{1}{\alpha+1}}~
\cosh^{\frac{2}{\alpha+1}} \left( \xi (\alpha+1) \Delta \varphi   \right) \nonumber \\
p_{\varphi} & = & \left( \frac{B}{A + 1}
\right)^{\frac{1}{\alpha+1}}~ \left[ A~\cosh^{\frac{2}{\alpha+1}}
\left( \xi (\alpha+1)  \Delta \varphi \right) -
\frac{A+1}{\cosh^{\frac{2 \alpha}{\alpha+1}} \left( \xi (\alpha+1) \Delta \varphi \right)} \right] \nonumber \\
\omega_{\varphi} & = & -~\frac{1 - A~\sinh^2 \left( \xi (\alpha+1)
\Delta \varphi \right)}{\cosh^2 \left(\xi (\alpha+1) \Delta \varphi
\right)} ~~.
\end{eqnarray}
For a flat universe (k = 0), using Eq. (3) , (12) and (13) we have
\begin{equation}
\Delta \varphi = \varphi -
\varphi_0=\pm\frac{2}{\sqrt{(D-1)(1+A)(1+\alpha)^2}}\sinh^{-1}\left[\frac{1}{A}\frac{a_{0}^{(D-1)(A+1)(\alpha+1)}}{a^{0.5(D-1)(A+1)(\alpha+1)}}\right].
\end{equation}
At last, we obtain the potential which has a form as
\begin{eqnarray}
U ( \varphi)= \frac{1}{2} \left( \frac{B}{A + 1}
\right)^{\frac{1}{\alpha+1}}\left[ \frac{1 + A}{\cosh^{\frac{2
\alpha}{\alpha+1}} \left( \xi (\alpha+1) \Delta \varphi \right)} +
(1-A) \cosh^{\frac{2}{\alpha+1}}\left( \xi(\alpha+1)\Delta \varphi
\right)\right].
\end{eqnarray}
Equations (17) are dynamic in term of time. In following , to get an
explicit form of the energy density and pressure according to the
scalar field, we assume a phenomenologically reliable power law
expansion of the scale factor $a(t)$ [14,23] as, $a(t) =
((t-t_0)^2+\frac{t_0}{1-\beta})^{\frac{1}{1-\beta}}$ so that, for
$\beta < 1$ we get accelerated expansion of the Universe thus
satisfying the observational constrains. In following ,we use this
setup for consideration of cosmological aspects of MCG model.\\
In this section, we considered the realization of the generalized
Chaplygin gas as a scalar field with a nontrivial potential.
However, this model construction cannot realize the background
equation of state across -1 due to a proof of No-Go theorem. This
No-Go theorem states that, any dark energy models realized by a
single scalar field or single perfect fluid within standard Einstein
gravity is not allowed to give rise to equation of state across -1,
otherwise the model suffers severe gradient instability at
perturbation level(for the detailed see[24]). In order to break the
No-Go theorem consistently, there are two simple mechanisms, which
can accommodate with the present study very well. One mechanism is
to introduce a higher derivative term nonlinearly, such as the
string theory inspired dark energy model proposed in [25]. The other
mechanism is to take the spinor field instead such as proposed in
[26]. Both two models are able to give rise to the generalized
Chaplygin gas behavior. Especially, the explicit realization of
Chaplygin gas was already discussed in the second mechanism. This
means, as authors proposed in [26] a model could constructed by
Spinor Quintom which combines the feature of a Chaplygin gas. The
generic expression of the potential is given by
$U=\sqrt[1+\gamma]{f(\varphi\bar{\varphi})+B}$ where
$f(\varphi\bar{\varphi})$ is an arbitrary function of
$\varphi\bar{\varphi}$. By Choosing $f(\varphi\bar{\varphi})$ to be
$f(\varphi\bar{\varphi}) = U_0(f(\varphi\bar{\varphi})- b)^2$ ,
where $U_0$, b, c are undetermined parameters, the crossing over -1
takes place when $\varphi\bar{\varphi} = b$ (for the detailed
see[26]).\\
\section{General Dark Energy Models}
In present work, we consider a MCG model as a candidate for dark
energy, but according to [28], there are many models that explain
dark energy. Hence, In this section for a eligible review, in
summary we point out them.\\
As we know several observational data (SNe Ia, CMB, LSS, BAO, WMAP,
SDSS) imply that the expansion of the universe is accelerating at
the present era. For explain this result, there are two
representative categories: 1. To add ``dark energy'' in the
right-hand side of the Einstein equation in general relativity. 2.
To modify the left-hand side of the Einstein equation, known as a
modified gravitational theory, (for example $F(R)$ gravity). In this
regards, the $\Lambda$ cold dark matter ($\Lambda$CDM) model has
been studied for playing a role of dark energy in framework of
general relativity. However, the theoretical derivation of the
cosmological constant $\Lambda$ has not been sorted out. In other
hand, other different models for dark energy without the
cosmological constant has been offered. For example, a canonical
scalar field, a non-canonical scalar field called as phantom,
tachyon scalar field and holographic dark energy.\\
Most important parameter to describe the aspects of dark energy
models is the equation of state (EoS)
$w_{\mathrm{DE}}=\frac{P}{\rho}$. In the framework of
Friedmann-Lema\^{i}tre-Robertson-Walker (FLRW) cosmology, there are
two ways to describe dark energy models. One is a fluid and the
other is to describe the action of a scalar field theory. In the
fluid proposal, we state the pressure as a function of energy
density $\rho$. But in the scalar field theory we induce the phrases
of the energy density and pressure of the scalar field from the
action of theory. In both proposals, we can imply the gravitational
field equations, this means that a cosmological model may be
described equally by different model illustrations.\\
In this article, we explicitly show that a cosmology with a fluid
description, also to describe in form of a scalar field theory. In
other words, the main goal of this article is to represent that one
dark energy model can exhibited as other dark energy models,
therefore, such a outcome unified illustration of dark energy models
might applied to any particular cosmology. Moreover, one can show
that degeneration among parameter of models can be removed by
precise data analysis of large data.\\
However,in following, we study a description of dark energy
universe. In this regards, we introduce some types of the
finite-time future singularities as well as the energy conditions in
the MCG model.
\section{Rip Singularity Scenario}
In this section, we will consider the modified Chaplygin gas model
which contain finite-time, future singularities. We know that
depending on energy density and pressure conditions of model such
singularities can lead to different ways. We classify these
situations in the following items[29,30]:
\begin{itemize}
\item  Type I (``Big Rip'') : For $t \to t_s$, $a \to \infty$,
$\rho \to \infty$ and $|p| \to \infty$
\item  Type II (``sudden'') : For $t \to t_s$, $a \to a_s$,
$\rho \to \rho_s$ and $|p| \to \infty$
\item  Type III : For $t \to t_s$, $a \to a_s$,
$\rho \to \infty$ and $|p| \to \infty$
\item  Type IV : For $t \to t_s$, $a \to a_s$,
$\rho \to 0$, $|p| \to 0$
\end{itemize}
where $t_s$, $a_s$ and $\rho_s$ are constants with $a_s\neq 0$. The
type I, known as the Big Rip singularity which appears for a phantom
equation of state: $w<-1$. The type II point out the sudden future
singularity which $a$ and $\rho$ be finite but $p$ diverges.\\
In figures (1-4), we show variation scalar field $\varphi$ , energy
density $\rho$ , pressure $p$ and equation of  state $\omega$ by
choosing arbitrary parameter of space as :  $A=0.66,~B=0.1
,~\alpha=0.5 ~
$and$ ~\beta=-4,~t_0=1 ~$in$~\Delta\varphi<0 $, for MCG model.\\
\begin{figure}[htp]
\includegraphics{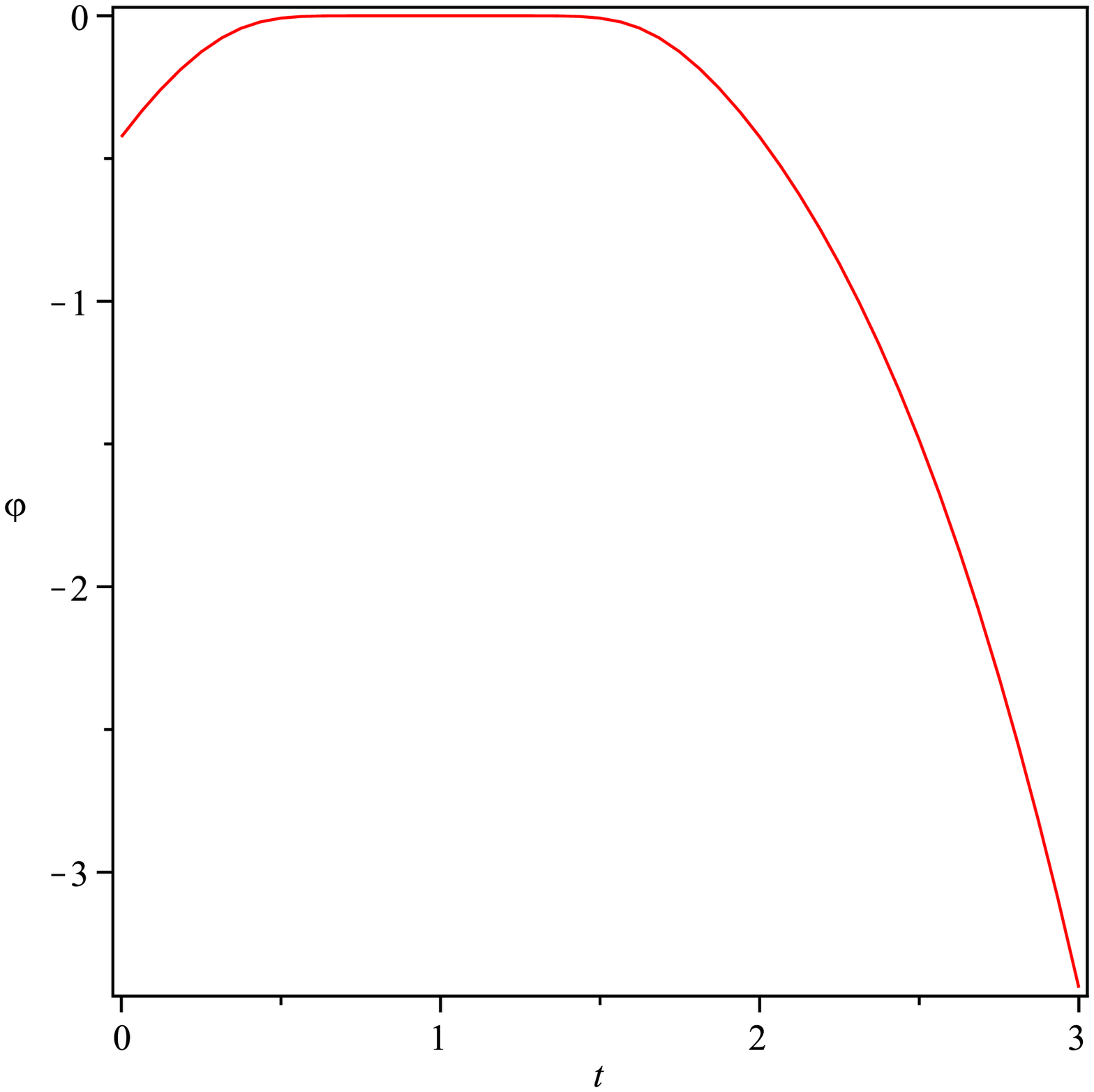} \vspace{7 cm}
 \caption{\small {Variation of the scalar field $\varphi(t)$. }
 \hspace{11cm}}
 \vspace{10cm}

\includegraphics{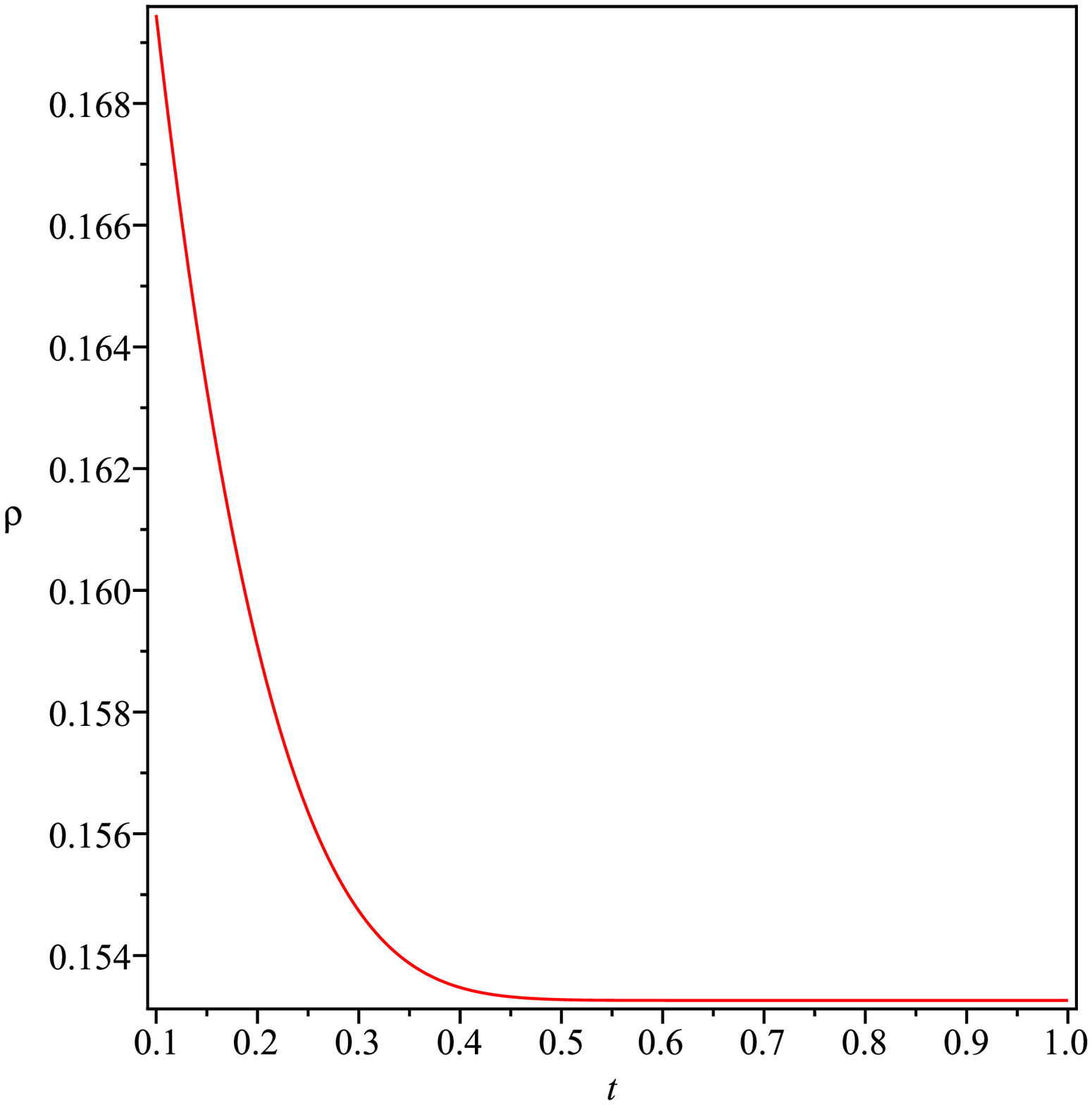} \vspace{-11 cm}  \caption{\small {Variation of the
energy density $\rho(t)$.}\hspace{-12cm}}
  \vspace{10cm}

\includegraphics{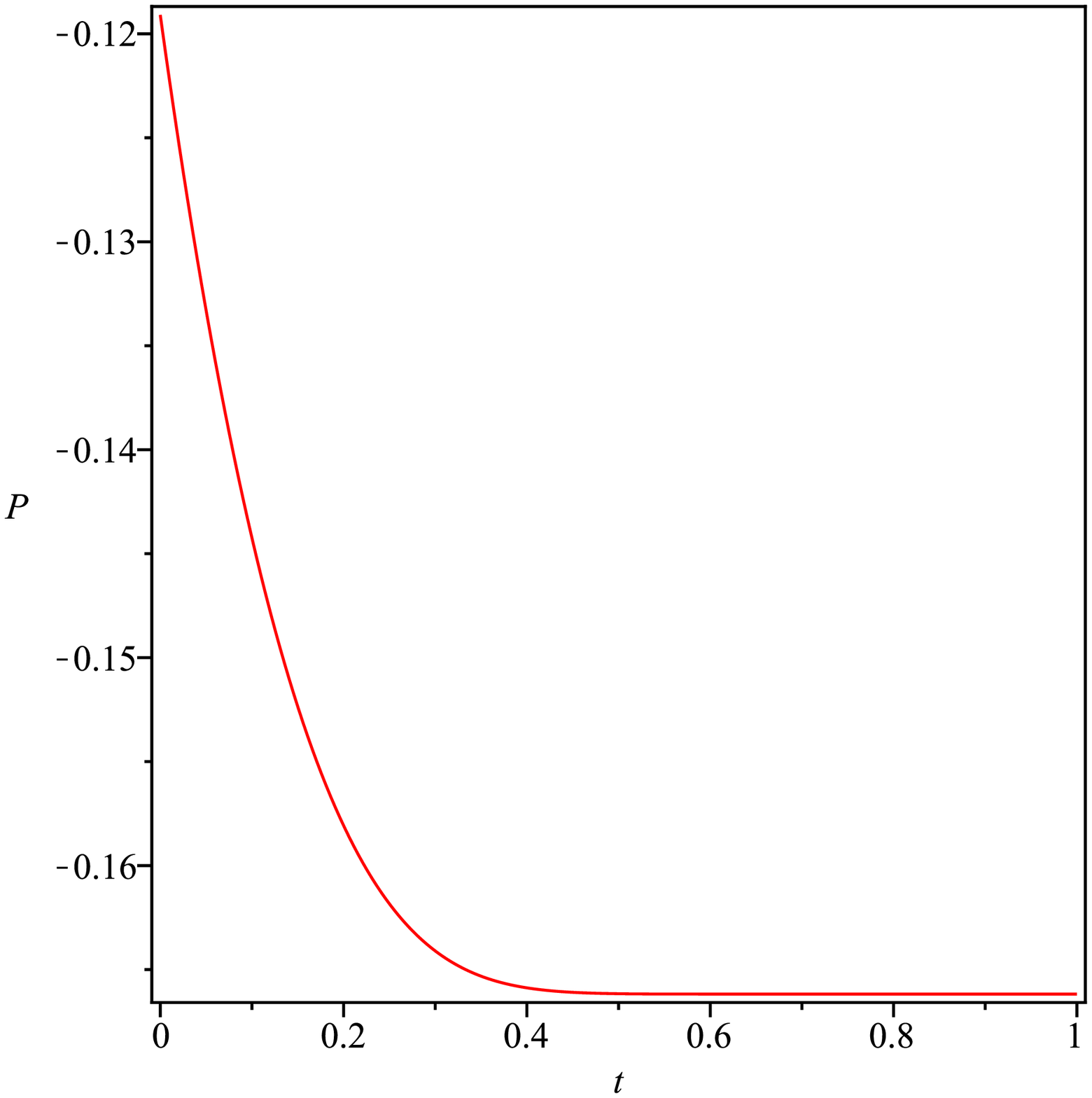}

\vspace{1 cm}
 \caption{\small {Variation of the pressure $p(t)$. }\hspace{11cm}}

\begin{center}
\includegraphics{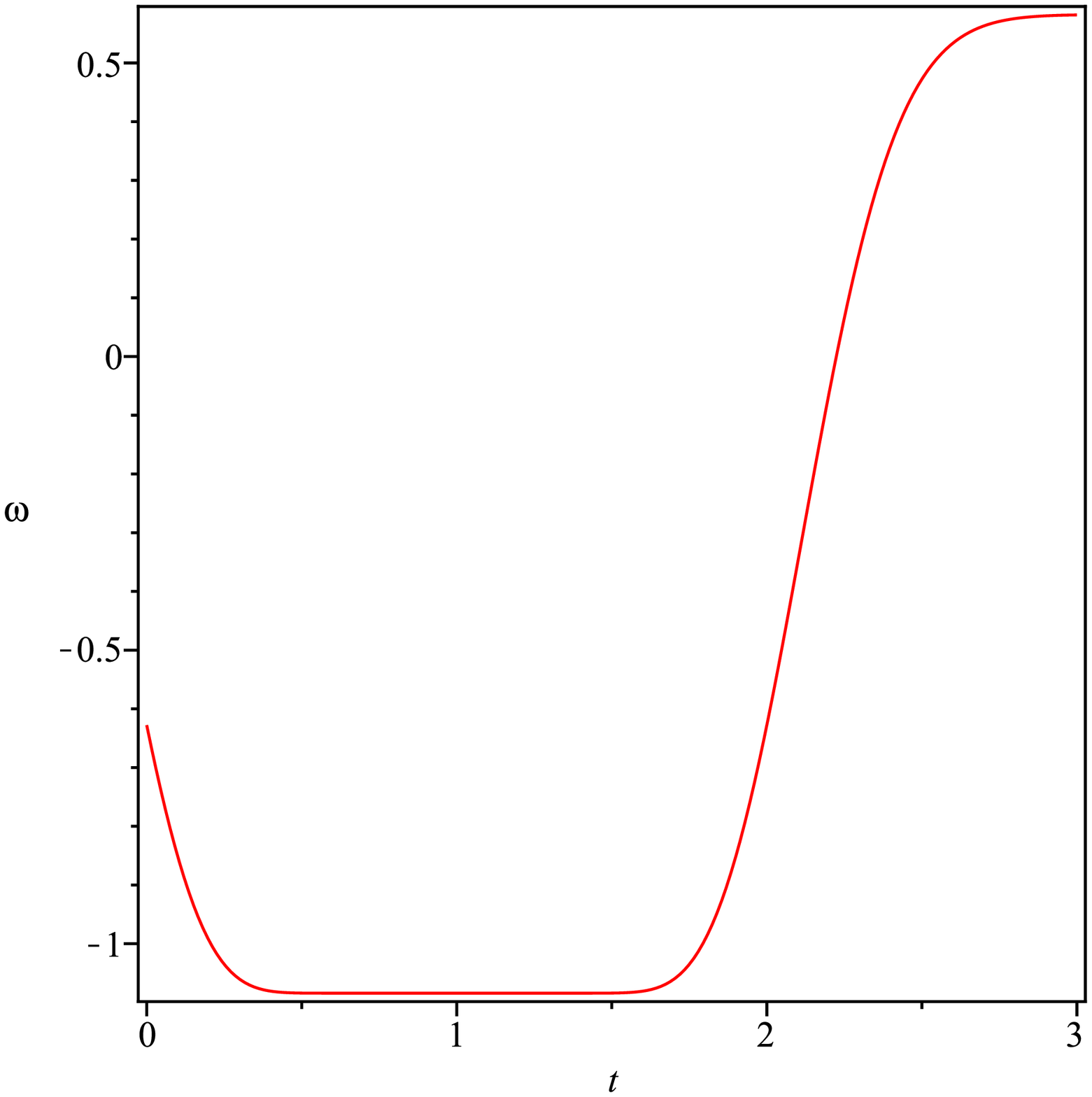}
\end{center}
\vspace{-2.5 cm}
 \caption{\small {Variation of the equation of  state $\omega(t)$.}\hspace{-12cm}}
 \end{figure}
If we choose different values for parameters of space of relations,
we can obtain different evolution of universe, these results
summarize in table (1). First, we consider the case of the Big Rip
singularity, Our calculation implies a Big Rip singularity occurs
just in $\beta>3$. If $\beta$ parameter be less 1, this type of
singularity never occurs. Second, as we shown in table (1) just type
IV singularity can appears in this setup of model and other types of
singularity can not produce.

\begin{table}[htp]
\caption{Summary of the behavior of MCG model depending specific
value of parameters. }
\begin{center}
\begin{tabular}
{llcccc} \hline \hline Type & $\alpha$ & $A$ & $B$ & $\beta$ &
$Appear~ in~ MCG$
\\[0mm]
\hline Type~I (``Big Rip") & $0\leq\alpha\leq1$ & $any$ & $B>0$ &
$>3$ & $Yes$
\\[0mm]
Type~II (``sudden") & $0\leq\alpha\leq1$ & $any$ & $B>0$ & $<1$ &
$No$
\\[0mm]
Type~III & $0\leq\alpha\leq1$ & $any$ & $B>0$ & $<1$ & $No$
\\[0mm]
Type~IV & $0.01$ & $0.99$ & $0.01$ & $-0.5$ & $Yes$
\\[1mm]
\hline \hline
\end{tabular}
\end{center}
\label{tb:table1}
\end{table}
\section{Bouncing Universe}
An interesting solution of the singularity problem of the standard
Big Bang cosmology known as Bouncing Universe. A bouncing universe
model has an initial narrow state by a non-zero minimal radius and
then develop to an expanding phase [14]. For a successful bouncing
model in the standard cosmology, the null energy condition is broken
for a period of time near the bouncing point. Moreover, for the
universe going into the hot Big Bang era after the bouncing, the
equation of state parameter of the universe should crossing from
$\omega<-1$ to $\omega>-1$.\\
In following, we study necessary conditions needed for a successful
bounce in a model of universe with modified Chaplygin gas. We know
that in the contracting phase, the scale factor $a(t)$ is
decreasing, this means, $\dot a(t)<0$, and in the expanding phase,
scale factor $\dot a(t)>0$. Finally in the bouncing point, $\dot
a(t)=0$, and near this point $\ddot a(t)>0$ for a period of time. In
other view, in the bouncing cosmology the Hubble parameter $H$
passes across zero
($H=0$) from $H<0$ to $H>0$.\\
Before bouncing point, we see that $\omega<-1$ and after the bounce,
the universe requires to enter into the hot Big Bang era, else the
universe filled with the matter by an equation of state $\omega<-1$
and occurs the big Rip singularity[31]. We can see from figure (5,6)
that in our setup a bouncing happens with the Hubble parameter $H$
going across zero and a minimal non zero scale factor $a$. At the
bouncing point $\omega$ has a finite negative value. So it is
possible to
realize bouncing solutions in a Chaplygin gas dark energy model.\\
\begin{figure}[htp]
\begin{center}\includegraphics{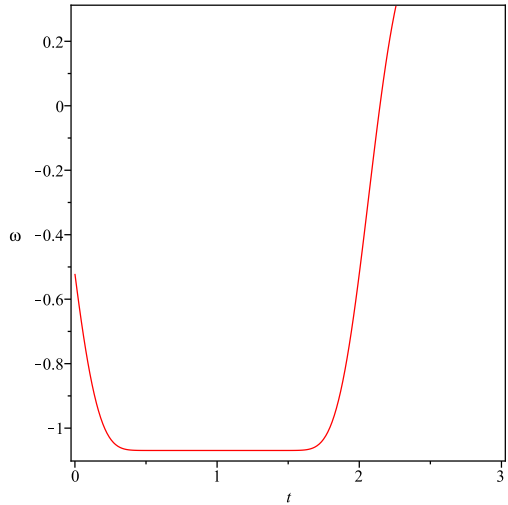} \vspace{6cm}
\end{center}
\caption{\small {Crossing of the phantom divide line by equation of
state parameter in a Chaplygin gas dark energy model for $A=0.62 ,
B=0.001 , \alpha=0.6 ~ and ~\beta=-3, t_0=1 $.}}
\end{figure}
\begin{figure}[htp]
\begin{center}\includegraphics{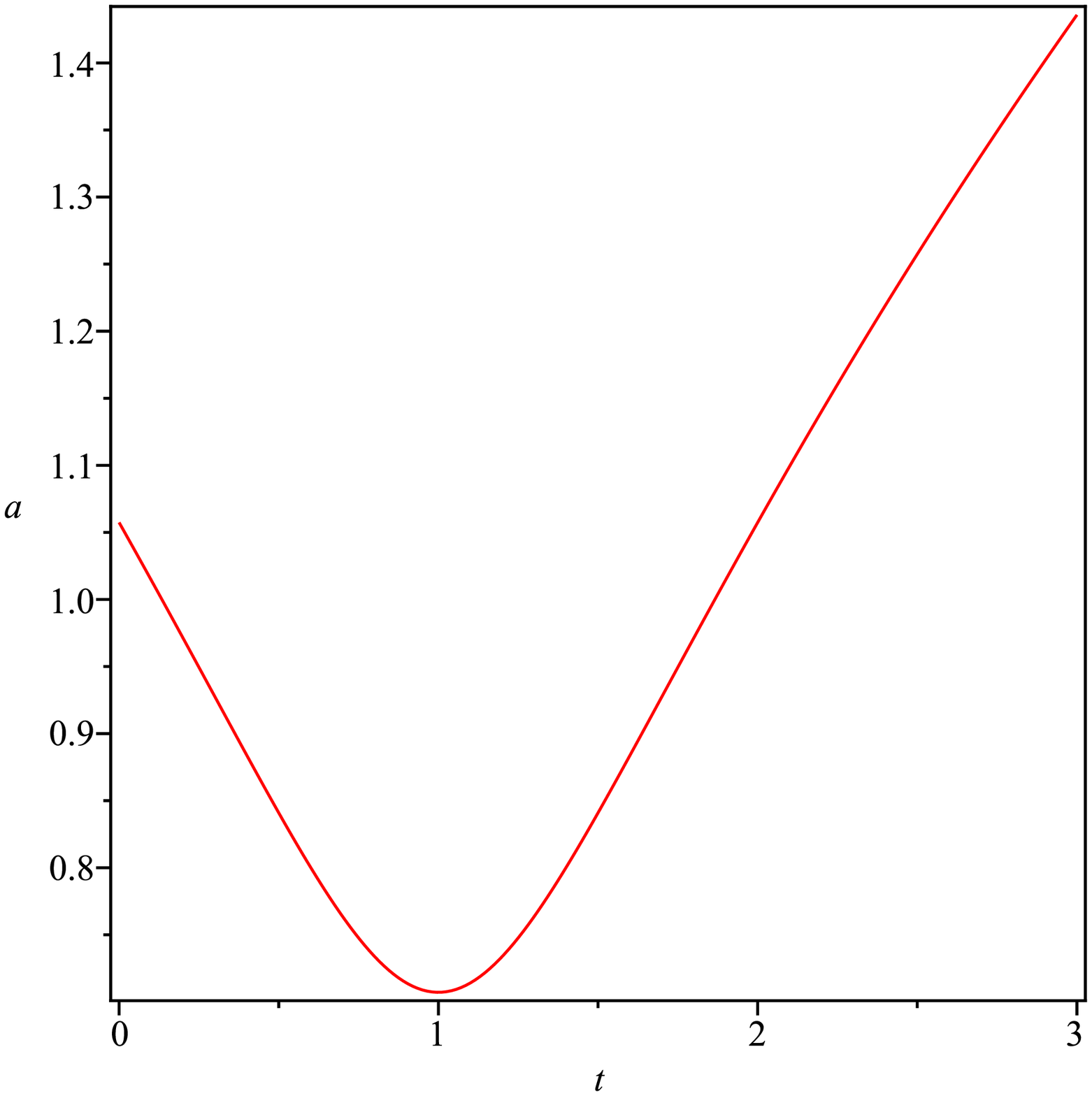} \vspace{5.5cm}\includegraphics{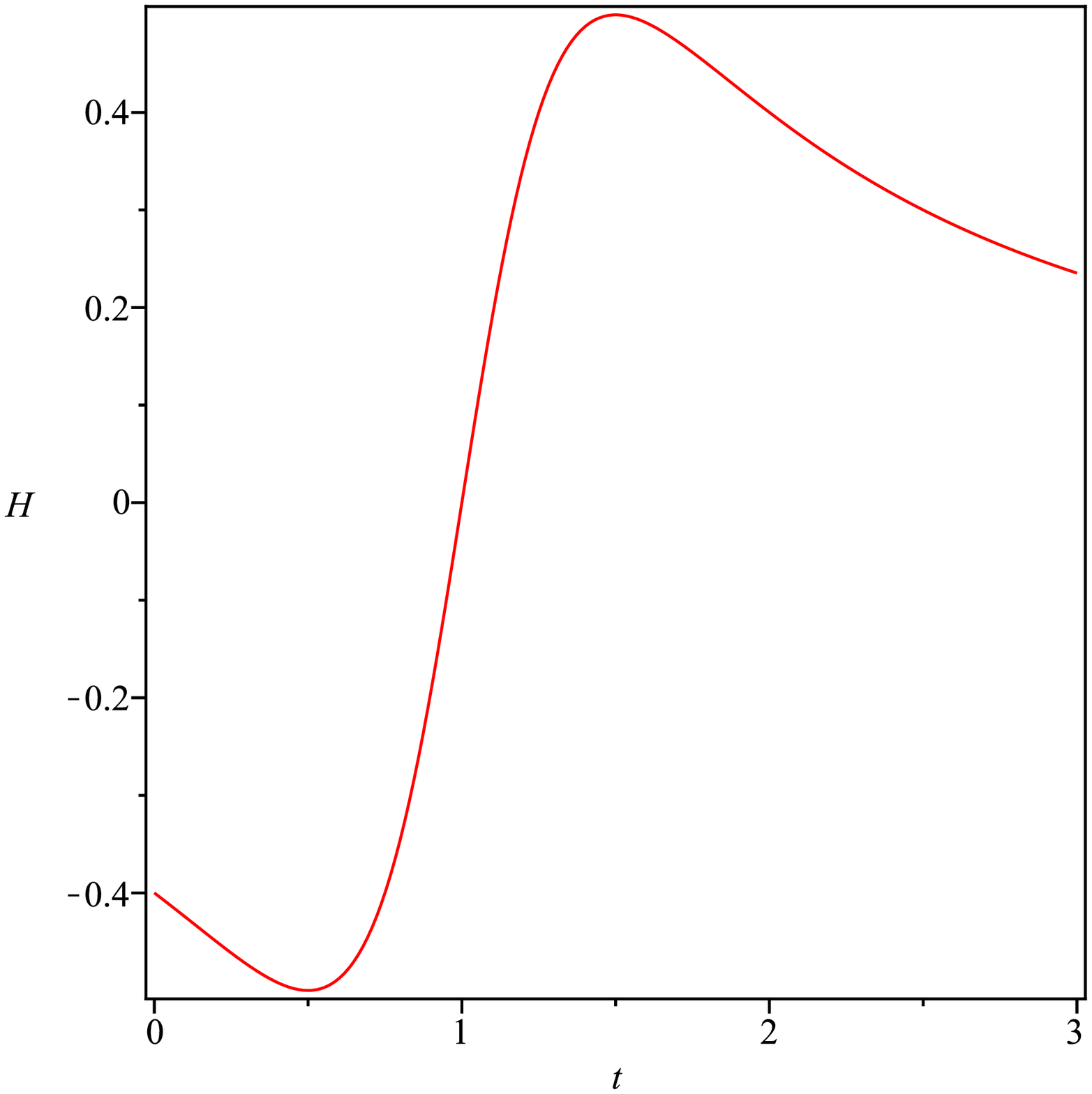}
\end{center}
\caption{\small {Variation of the scale factor $a$ relative to
cosmic time $t$ (left), and Hubble parameter $H$ (right) for $t_0=1
~and~ \beta=-3$. }}
\end{figure}
Therefore, we shown that by a suitable choose of parameter space,
this model can explain a Bouncing solution. We emphasize range of
parameter that used in numerical analyze adjust with observational
constraints [20,27,32]. Overall, observational signatures in
cosmological surveys related to the cosmological perturbation theory
widely studied in the literature. For theoretical aspect, the
corresponding perturbation theory within bouncing cosmology was
developed in [33,34,35]. And for observational aspect, the
comparison of bouncing cosmology with data can be found in [36,37].
\section{Summary}
In the present article, we have considered finite-time future
singularities in modified Chaplygin Gas model. We have shown this
model realizing a crossing of the phantom divide line. Also it has
been shown that some types of singularity may appear in MCG model.
This means, in this framework, there is the possibility of a Rip
singularity by suitable tuning in the parameters. Moreover, our
results show that such a model universe with Chaplygin Gas as dark
energy component avoids the problem of the Big Bang singularity with
a bouncing scenario.\\

\end{document}